\documentclass{article}
\usepackage{amsmath,amsthm,amssymb,a4wide,scalerel,url,xcolor,tikz,algorithm,caption,algpseudocode,enumitem}
\usetikzlibrary{arrows.meta}
\usepackage[symbol]{footmisc}

\algtext*{EndFor}{} \algtext*{EndWhile}{} \algtext*{EndIf}{}
\tikzset{pt/.style = {circle,fill,inner sep=2pt}}

\newtheorem*{theorem*}{Theorem} \newtheorem{theorem}{Theorem}[section] \newtheorem{proposition}[theorem]{Proposition} 
\newtheorem{lemma}[theorem]{Lemma} \newtheorem*{lemma*}{Lemma} \newtheorem{corollary}[theorem]{Corollary}
\theoremstyle{definition} \newtheorem*{definition}{Definition} \newtheorem*{question}{Question}

\newcommand{\eps}{\varepsilon} \newcommand{\emptystr}{\epsilon}
\newcommand{\Pon}{P_{\textnormal{on}}} \newcommand{\Mon}{{M}_{\textnormal{on}}} \newcommand{\tMon}{{\widetilde M}_{\textnormal{on}}} 
\newcommand{\smalllt}{\hspace{-1pt}\scaleobj{0.8}{<}\hspace{-.5pt}} \newcommand{\omgle}{\omega_{\scaleobj{0.8}{\le}n}} \newcommand{\omglt}{\omega_{\smalllt n}}
\newcommand{\goodn}{n:\omglt \bar\omega_n \not\in}
\DeclareMathOperator{\K}{K} \renewcommand{\mid}{\,|\,}

\title{Solomonoff induction of selected bits and online classification}
\author{Bruno Bauwens, German Kuznetsov\footnote{
  Both authors work in National Research University Higher School of Economics, Faculty of Computer Science, Moscow, Russia.
  This work/article is an output of a research project (HSE-BR-2025-024) implemented as part of the Basic Research Program at HSE University.
  \newline Results were announced in the International Conference Dedicated to the 125th Anniversary of P.S. Novikov~\cite{NovikovTalkBruno2026}.
(August 24–28, 2026, Steklov Mathematical Institute of RAS, 8 Gubkina St., Moscow), 
}}
\date{\today}
\begin{document}
\maketitle

\begin{abstract}
  Solomonoff induction is a general solution for sequence prediction that started the field of algorithmic information theory around 1960. 
  Let $M$ be the universal semimeasure. 
  After observing a string $\omega_1 \cdots \omega_{n-1}$, it predicts the next bit $\omega_n$ to be $b$ with conditional probability proportional to $M(\omega_1 \cdots \omega_{n-1} b)$.  
  Thus the error of this probability is
  \[
    e_\omega(n) = 1 - \frac {M(\omega_1 \cdots \omega_{n-1}\omega_n)} {M(\omega_1 \cdots \omega_{n-1} 0) + M(\omega_{1} \cdots \omega_{n-1} 1)}. 
  \]
  If $\omega$ a computable infinite sequence with a self-delimiting description of at most $k$ bits, then $\sum_n e_\omega(n) \le k\ln 2+O(1)$. 
  
  Lattimore, Hutter and Gavane consider the problem of predicting computable patterns in a sequence, 
  for example, the pattern of repeated bits $\omega = 00\,11\,00\,00\,11\,00\cdots$
  They show that the error probabilities of the bits fixed by a computable pattern have limit zero, for example, for repeated bits, one has $\lim_n e_\omega(2n) = 0$. 

  We show that for some computable pattern, the sum of $e_{\omega}(n)$ over positions $n$ from the pattern is not bounded. 
  However, the sum of squares $(e_\omega(n))^2$ is bounded exponentially in the program length of the pattern, 
  It is also shown that bits of $\omega$ that are fixed by a computable classification function $f:\{0,1\}^* \rightarrow \{0,1\}$, 
  converges faster: the sum of $(e_\omega(n))^2$ is bounded by a polynomial in the program length 
  for programs whose runtime is bounded by any fixed computable time bound. 
\end{abstract}

\section{Introduction}

Many machinelearning tasks can be viewed as sequence prediction: given bits $\omega_1, \ldots, \omega_{n-1}$, what is the probability of the next bit $\omega_n$ being $0$? 
Solomonoff provided a simple mathematical solution for this problem based on algorithmic probability~$M$ defined below~\cite{Solomonoff1960}. 
In this way, he united various philosophical approaches to the problem of scientific inference and the foundations of probability, see~\cite[chapter 5]{LiVitanyiForthEdition}. 

Often, we are only interested in the errors of some of the bits. 
For example, ``online classification'' can be viewed as ``sequence prediction'' by building a sequence that alternatingly consists of input features and labels. 
In this case, we do not care about predictions of the input features, but only about their labels. 
In this paper, we determine convergence rates for predicting computable patterns in sequences and in particular for deterministic online classification rules.

\begin{definition}\label{def:semimeasure}
  A  {\em semimeasure} is a function $P:\{0,1\}^* \rightarrow \mathbb R_{\ge 0}$ that satisfies 
  $P(\text{empty str}) \le 1$ and $P(x) \ge P(x0) + P(x1)$, 
  where $xb$ denotes the concatenation of string $x$ and bit $b$. 
\end{definition}

Among the lower semicomputable semimeasures, there exists a semimeasure $M$ that is {\em maximal},  
which means that for each such semimeasure $P$, one has $\exists c \forall x[P(x) \le cM(x)]$. 
Note that all maximal such semimeasures are equal up to a constant factor. 
Let~$M$ be any such maximal semimeasure. 

Solomonoff's solution for sequence prediction given $\omglt = \omega_1 \cdots \omega_{n-1}$ is to predict $\omega_n = b$ with probability\footnote{ 
  \label{footnote:conditioning}
  In the context of convergence of subpatterns, it is important that the denominator 
  is normalized and equals $M(\omglt0) + M(\omglt1)$ instead of the commonly used $M(\omglt)$. 
  Indeed, in \cite[theorem 12]{Lattimore2011selectedBits} the case of repeated bits is studied. 
  It is shown that there exist $\omega \in \{0,1\}^\infty$ with $\omega_{2n-1} = \omega_{2n}$ for all $n$ such that
  \[
    \liminf_{n \rightarrow \infty} \frac {M(\omega_{{<}2n} 0) + M(\omega_{{<}2n} 1)} {M(\omega_{{<}2n})} < 1. 
  \]
  However, the definition of algorithmic probability $M$ is different in \cite{Lattimore2011selectedBits}. 
  We think this also holds for certain maximal semimeasures $M$, (but not for all). 
  }
\[
  M(b \mid \omglt) = \frac { M(\omglt b)} {M(\omglt 0) + M(\omglt 1) }. 
\]

\subsection*{Computable patterns}

It is well known that Solomonoff induction quickly converges on computable sequences, (even for the unnormalized conditionals). 
Indeed, if $k$ is the minimal prefix complexity of a program that computes $\omega$, then by~\cite[theorem 3.8.9]{Hutter2024book}\footnote{
  We switched from monotone to the larger prefix complexity because it is better known and we use it elsewhere in this paper. 
  The absolute-error bound follows from
  \cite[eq (8) section 3]{Hutter2003convergence}, after applying $1-p \le \ln \tfrac 1 p$. 
}, 
\[
  \sum_n 1 - \frac {M(\omgle)} {M(\omglt)} \le k \ln 2 + O(1).
\]

\noindent
The following question was raised in~\cite[problem 2.9 p63]{Hutter2005universal} and~\cite{Hutter2009openProblems}. 

\begin{question} 
  Does conditional probability $1 - M(\omgle)/M(\omglt)$ decrease rapidly to zero on the bits that are fixed by some computable pattern? 
  For example, if each even bit equals the preceding odd bit, does the sum remain bounded?
\end{question}

\noindent
In~\cite[theorem 11, p269]{Lattimore2011selectedBits} 
it was shown that $M(\omega_n \mid \omglt)$ does converge to 1 for those bits $\omega_n$ that are fixed by a computable pattern. 
Unfortunately, the result depends on a proof cited as private communication~\cite{privateLetterAlgorithmicProbabilities}, and we did not find a public version. 
For this reason, also no quantitative bounds were found. 

Our first result provides upper bounds on the convergence rates for $L_\gamma$-losses with $\gamma>1$ and $\gamma = 1$. 
For a bit $b$, let $\bar b$ be its negation.

\begin{theorem}\label{th:main_exponential}
  Fix $\gamma > 1$. 
  Let $T \subseteq \{0,1\}^*$ be decided by a program of length~$k$. 
  If all prefixes of $\omega \in \{0,1\}^\infty$ are in $T$, then
  \[
    \sum_{\goodn T} \left(M(\bar \omega_n \mid \omglt)\right)^{\gamma} \le O(k^{2\gamma} 2^{\gamma k}) 
  \]
  and 
  \[
    \sum_{\substack{\goodn T\\n \le N}} M(\bar \omega_n \mid \omglt) \le O(k^22^k \log^2 N \log^2 \log N).
  \]
\end{theorem}

\noindent
The above bounds are exponential in~$k$ and we do not know if polynomial bounds hold. 
Our proof relies on a game that was earlier defined by Mikhail Andreev and Akim Kumok~\cite{Andreev2016sum}, 
but they mostly focus on the opposite player. 

\bigskip
\noindent
The following result shows that for $\gamma = 1$, the errors diverge slowly. 
A  {\em tree} $T$ is a subset of $\{0,1\}^*$ that is closed under taking prefixes.  

\begin{theorem}\label{th:main_L1}
  There exists a computable leafless tree $T \subseteq \{0,1\}^*$ and infinitely many strings $w$ in $T$ such that
  \[
    \sum_{\substack{n:w_{\smalllt n}\bar w_n \not\in T\\n \le |w|}} M(\bar w_n \mid w_{\smalllt n}) \ge \Omega\left(\tfrac {\log |w|} {\log^2 \log |w|}\right).
  \]
\end{theorem}

\noindent
This theorem follows from the proof of a result by Mikhail Andreev and Akim Kumok~\cite{Andreev2016sum}. 


\subsection*{Online classification}

Solomonoff proved that for random sequences, algorithmic probability rapidly converges to conditional measure~\cite{Solomonoff1978}.
If $P$ is a measure computed by a self-delimiting program of length $k$, then for an infinite sequence $\omega$ generated from $P$,
one has
\[
  \underset{\omega \sim P}{\mathbb E} \sum_{n=1}^\infty \big(P(0 \mid \omglt) - M(0 \mid \omglt) \big)^2 \le k \ln \sqrt 2 + O(1). 
\]
Now we consider the problem of online classification of strings of arbitrary length. 
This can be cast as sequence prediction using any prefix-free encoding of $\langle x \rangle$ of strings~$x$, 
where $\omega = \langle x_1 \rangle f(x_1) \langle x_2 \rangle f(x_2) \cdots$ with $x_i \in \{0,1\}^*$. 
Now we are interested in the sums of errors only for those bits $\omega_n$ that correspond to values $f(x_i)$. 

Can we obtain a small mistake bound if there exists a simple classifier? 
Unfortunately, the above bound is only small when the combined complexity of input generation and true classifier are small. 

For example, imagine one wants to train a classifier for discriminating black-white movies from color movies in mp4 files. 
Such a classifier should be relatively easy to learn, however, any realistic measure $P$ over movies might be quite complex. 
In fact, it is not even clear what ``realistic measure'' means.


Our next result states that for online classifiers, the L$\gamma$-norm of the probability error is polynomial in the Kolmogorov complexity of the classifier. 
Thus, this is exponentially better than for general patterns. 

To define $M(b \mid L)$ for a list $L$ of strings, 
fix a computable prefix-free encoding of strings,\footnote{
  We assume that the question whether $y$ is a prefix of the encoding of some string is decidable.
  When only using lists $L$ in which the words at even positions are single bits, then it is not needed to apply an encoding on these bits. 
}
and encode $L$ by concatenating the prefix-free encodings of its elements.

\begin{theorem}\label{th:classifiers_upper}
  Fix $\gamma>1$ and a computable time bound~$t$. 
  If $f: \{0,1\}^* \rightarrow \{0,1\}$ is computed by a $k$-bit program in time $t(k,|z|)$ on input~$z$, then for $x_1, x_2, \ldots \in \{0,1\}^*$, 
  \[
    \sum_{n=1}^\infty \big(1 - M(f(x_n) \mid h_{n}, x_n)\big)^{\gamma} \le O(k^{1+\gamma} \log^{2\gamma} k), 
  \]
  where $h_n = [x_1, f(x_1), \ldots, x_{n-1}, f(x_{n-1})]$.  
\end{theorem}

\noindent
In this result, the time bound affects the constant in the big-O. 
A similar statement can be proven for functions $f$ that output strings of constant length, but we do not know how to prove it for arbitrary output. 

\begin{corollary}\label{cor:classifiers_lower}
  There exists a computable classifier $f$ and infinitely many $N$ for which there exist strings $x_1, x_2, \ldots, x_N \in \{0,1\}^*$ such that 
  \[
    \sum_{n=1}^N M(1-f(x_n) \mid h_n, x_n) \ge \Omega \left(\frac {\log N} {\log^2 \log N}\right),
  \]
  where $h_n = [x_1, f(x_1), \ldots, x_{n-1}, f(x_{n-1})]$.  
\end{corollary}

In fact, the above theorem is true for the constant $0$ classifier. 
In appendix \ref{sec:online} we discuss online algorithmic probability, which was defined in~\cite{onlineComplexity} 
(and used in~\cite{Hutter2005universal,Hutter2024book} for reinforcement learning). 
If in the above sum, conditional algorithmic probability were replaced by online probability, then the sum is always bounded by a constant. 
Thus with online algorithmic probability one can obtain better bounds for the absolute loss.

\section{Games}

Note that we may always assume that sets $T$ in the theorem are closed under prefixes, 
because dropping strings that do not contain all their prefixes does not change the set of infinite $\omega$ that are applicable to the theorems, neither does it affect the sums.

\subsection*{The flow-flow game}

{\em The flow-flow game.} The game has parameters $s, \gamma$  and is played on $K$ trees $T_1, \ldots, T_K \subseteq \{0,1\}^*$, 
which are known to both players. 
Two players, called Alice and Bob, each maintain their own continuous semimeasure, denoted $A$ and~$B$. 
Initially, $A = B = 0$. 
They alternate turns and at their turn, they can increase finitely many values of their semimeasure, 
provided functions $A$ and $B$ remain semimeasures.

{\em Alice wins} immediately if after some move of Bob, there exists an infinite branch $\omega$ of some tree $T_j$ such that
\[
  \sum_{\goodn T_j} \left(\frac {A(\omglt \bar \omega_n)} {B(\omglt0) + B(\omglt1)}\right)^{\gamma} \;>\;s, 
\]
where in the sum, $0/0 = 0$ is assumed. {\em Bob wins} if he can reply indefinitely without Alice winning.

\medskip
\noindent
{\em Example 1.} 
Consider the game with $s = \gamma = 1$ and the single tree $T = \{\text{empty str.}, 0, 00, 000, \ldots\}$.
Bob wins by concentrating all measure along the tree, thus setting $B(x) = 1$ for all $x \in T$. 

\medskip
\noindent
{\em Example 2.} 
Again, let $s = \gamma = 1$. Consider the game with an arbitrary single tree. 
If Alice evaluates a computable rational measure~$A$, then Bob has a computable winning strategy
by splitting his flow proportionally to $A$'s flow in nodes $x$ for which both children are in the tree, 
and otherwise, if $x$ has only 1 child, sending all flow in this child's direction. 

Indeed, at every round, the support of $A$ is finite, so we prove it by induction on the maximal length $d$ of the support. 
Suppose that the strategy works for some $d$ and all trees. Consider some $A$ supported on $(d+1)$-bit strings. 
If the root of the tree has 2 children, then the root does not contribute to the sum. 
Consider the 2 inductive applications corresponding to the children of the root. 	 
Compared to the original sum, both the numerators and denominators are multiplied by the same factors, thus Bob wins. 
Otherwise, if the root has 1 child $b$, then the root contributes an extra term $A(\bar b)$ to all sums. 
But by the inductive hypothesis, the strategy $B'(x) = B(bx)$ wins against strategy $A'(x) = A(bx)/A(b)$, 
meaning that in our game, the sum is bounded by $s\cdot A(b) = A(b)$. Thus in each branch, the sum is at most $s=1$.

\subsection*{Winning strategies and error bounds}

Strategies for Alice provide lower bounds in the main theorems. 
Strategies for Bob provide upper bounds. 
For the exponential upper bounds in theorem~\ref{th:main_exponential}, only the game with $K=1$ tree is needed. 
For a set $S \subseteq \{0,1\}^*$, let $\K(S)$ be the minimal prefix complexity of a program that decides~$S$, 
see~\cite{LiVitanyiForthEdition,bookShenVereshchagin} for background on algorithmic information theory.

\begin{lemma}\label{lem:game2thm_1tree}
  Fix $\gamma \ge 1$ and a constant~$s$. 
  Assume that for each tree $T$, Bob has a winning strategy in the game with 1 tree~$T$ and value~$s$, 
  and that these strategies are uniformly computable given~$T$. 
  Then for each computable $T$ and infinite branch~$\omega$ in~$T$, 
  \[
    \sum_{\goodn T} \big(M(\bar \omega_n \mid \omglt)\big)^{\gamma} \le O\big(2^{\gamma\K(T)}s\big). 
  \]
\end{lemma}

\noindent
The proof of the lemma uses the following result, see for example~\cite[cor 4.5.1, p307]{LiVitanyiForthEdition}. 
\begin{lemma}\label{lem:MdomP}
  For all lower semicomputable semimeasures $B$ and $x$,
  \[
    2^{-\K(B)}B(x) \le O(M(x)). 
  \]
  where $\K(B)$ is the minimal prefix complexity of a program that lower approximates~$B$. 
\end{lemma}

\begin{proof}[Proof of lemma~\ref{lem:game2thm_1tree}.] 
  Consider the game for a computable tree $T$. 
  Let Alice play a strategy in which she ignores $T$ and Bob's moves:
  she computes a lower approximation of~$M$ and in each round, sets the current values equal to~$A$. 

  Bob plays the winning strategy. 
  We obtain a semimeasure~$B$ by taking the limit of his semimeasure. 
  It satisfies
  \[
    \sum_{\goodn T} \left(\frac {M(\omglt \bar\omega_n)} {B(\omglt0) + B(\omglt1))} \right)^{\gamma} \le s. 
  \]
  Note that $\K(B) \le \K(T) + O(1)$, because $B$ is computed from $T$ by evaluating the strategy.
  Hence,  lemma~\ref{lem:MdomP} implies
  \[
    M(\bar \omega_n \mid \omglt) \le 2^{\K(T)}
    \frac {M(\omglt \bar \omega_n)} { B(\omglt 0) + B(\omglt 1)}
  \]
  The lemma finishes by raising to the power $\gamma$ and summing. 
\end{proof}

\begin{lemma}\label{lem:game2thm}
  Fix $\gamma \ge 1$, a computable function $s : \mathbb N \rightarrow \mathbb Q$, and a uniformly computable sequence of trees $T_1, T_2, \ldots$
  Suppose Bob wins each game with parameters $\gamma, s_\ell$ and trees $T_1, \ldots, T_{2^{2^\ell}}$. 
  Also suppose that the strategies are uniformly computable in~$\ell$. 
  Then, for each tree $T_j$ and each infinite branch $\omega$ in $T_j$, 
  \[
    \sum_{\goodn T_j} (M(\bar \omega_n \mid \omglt))^{\gamma} \le O(\ell^{2\gamma} s_\ell). 
  \]
\end{lemma}

\begin{proof}[Proof sketch.]
  The proof is similar as for lemma~\ref{lem:game2thm_1tree}, but with a different estimate for the complexity of Bob's semimeasure. 
  Now the trees and the strategies are uniformly computable in $\ell$, thus this complexity is bounded by the complexity of~$\ell$. 
  Hence, $\K(B_\ell) \le \K(\ell) + O(1) \le 2\log \ell + O(1)$. 
  After taking exponents and $\gamma$ powers, this corresponds to the factor~$O(\ell^{2\gamma})$ in the right side. 
\end{proof}

\medskip
\noindent
The second part of theorem~\ref{th:main_exponential} requires bounds on the sum for $N$-bit prefixes of~$\omega$ and   
$s$ is a function. Alice wins if after some move of Bob, there exists an infinite branch $\omega$ of some tree $T_j$ and an $N$ such that
\begin{equation}\tag{$*$}\label{eq:winning}
  \sum_{\substack{\goodn T_j\\n \le N}} \left(\frac {A(\omglt \bar \omega_n)} {B(\omglt0) + B(\omglt1)}\right)^{\gamma} \;>\;s_N. 
\end{equation}
It is easy to adapt the above lemmas when $s$ is a function instead of a constant. This was omitted for notational convenience.

\begin{lemma}\label{lem:game_lowerbound}
  Let tree $T$ and function $s$ be computable. 
  Assume that for all $N$,
  Alice has a strategy in which for all $N$ there exits an $N$-bit string $z$ in $T$ such that for infinitely many rounds,
  \begin{equation}\tag{$*$}\label{eq:winning}
    \sum_{\substack{z_{\smalllt n}\bar z_n \not\in T\\n \le N}} \left(\frac {A(z_{\smalllt n} \bar z_n)} {B(z_{\smalllt n}0) + B(z_{\smalllt n}1)}\right)^{\gamma} \;>\;s_N, 
  \end{equation}
  If these strategies are uniformly computable given $N$, 
  then for all $N$, there exists an $N$-bit $z$ in $T$ such that
  \[
    \sum_{n:z_{\smalllt n}\bar z_n \not\in T} (M(\bar z_n \mid z_{\smalllt n}))^{\gamma} \ge \Omega(s_N2^{-\gamma \K(N)}).
  \]
\end{lemma}

\begin{proof}
  Consider a game in which Bob plays blindly, and lower approximates a priori probability~$M$. 
  Alice plays the winning strategy. 
  Since Alice's semimeasure $A$ is lower semicomputable given~$N$, lemma~\ref{lem:MdomP} implies
  \[
    M(x) \ge \Omega (2^{-\K(N)}A(x)), 
  \]
  The proof finishes by substituting this, raising to the power $\gamma$ and summing. 
\end{proof}

\subsection*{The weight-flow game}

We now consider an equivalent game for the case of a single tree~$T$. 
The flow that leaves Alice's tree in a certain node is viewed as a weight. 

Formally, let $\partial T$ denote the set of all nodes that are not in~$T$ but whose parent is in~$T$. 
Thus $\partial T$ is a prefix-free set. Let $b$ always represent a bit. 
For all $xb \in \partial T$ (implying $x \in T$), we set $w(x) = A(xb)$. 
(If $x$ is a leaf, then also $x \bar b \in \partial T$. One may choose $w(x) = 0$, but this won't matter.)

This leads to an equivalent game, called the {\em weight-flow} game, 
where Alice controls weights instead of flows. 
We additionally allow weights on all the nodes, and it turns out that we need no tree $T$ at all. 

First we formally prove that this game is indeed harder. 
Afterwards, we prove that the harder game is equivalent to the flow-flow game with the tree for repeated bits.

\bigskip
\noindent
{\em The weight-flow game.} 
The game has parameters $\gamma$ and~$s$. 
Two players Alice and Bob alternate turns. 
As before, Bob maintains a semimeasure~$B$. Initially $B=0$. 
Alice maintains weights $w(x) \ge 0$ for all strings $x \in \{0,1\}^*$ such that 
\[
   \sum_{x \in \{0,1\}^*} w(x) \le 1. 
\]
Initially $w=0$. 
In each round Alice increase finitely many values of~$w$. 
Afterwards, Bob increases finitely many values of~$B$. 

{\em Alice wins} if after Bob's move, there exists an $\omega \in \{0,1\}^\infty$ with
\[
  \sum_{x \sqsubset \omega}\left( \frac {w(x)} {B(x)} \right)^{\gamma} > s,
\]
where $0/0 = 0$ is assumed. 
 {\em Bob wins} if he can reply indefinitely without Alice winning.

\newcommand{\Aff}{A_{\textnormal{ff}}} \newcommand{\Bwf}{B_{\textnormal{wf}}} \newcommand{\Bff}{B_{\textnormal{ff}}} 
\newcommand{\Bwfj}{B_{\textnormal{wf},j}} \newcommand{\Bffj}{B_{\textnormal{ff},j}}

\begin{lemma}\label{lem:weightflow2flowflow_Bob}
  If Bob has a winning strategy in the weight-flow game, 
  then for each single tree~$T$, Bob has a winning strategy in the flow-flow game with the same parameters $\gamma$ and~$s$. 
\end{lemma}

\begin{proof}
  The proof is a simple formal verification. 
  Let $\Aff$ be Alice's strategy in a flow-flow game with a single tree~$T$. 
  We need a winning strategy $\Bff$ for Bob in this game. 

  First $\Aff$ is transformed to a strategy $w$ for Alice in the weight-flow game. 
  Recall that $b$ is always a bit. 
  For each node $xb$ in $\partial T$ with $x\bar b \in T$, Alice sets $w(x) = \Aff(xb)$. She keeps $w(x)=0$ for all other~$x$.
 
  Bob plays his winning strategy $\Bwf$ against~$w$. 
  He obtains a strategy $\Bff$ by ``pushing $\Bwf$ one step forward'' 
  when the relevant direction in $T$ is unique. More precisely, $\Bff(\epsilon) = \Bwf(\epsilon)$
  \\-- $\Bff(xb) = 0$ if $xb \not\in T$, 
  \\-- $\Bff(xb) = \Bwf(xb)$ for nodes $xb \in T$ with a brother, thus if $x \bar b \in T$, 
  \\-- $\Bff(xb) = \Bwf(x)$ for nodes $xb \in T$ without brother. 

  Note that $\Bff(y) \ge \Bwf(y)$ for $y \in T$ and that $\Bff(y) > \Bwf(y)$ only happens if $y$ is in $T$ and has no brother in $T$. 
  This implies $\Bff(x) \ge \Bff(xb) + \Bff(x \bar b)$ for all $x$, because if $x\bar b \not\in T$, then $\Bff(p \bar b) = 0$. 

  In the sum that defines the winning condition, only values $\Bff(xb)$ are used for which $xb$ has no brother. Thus we have
  \[
    \frac {\Aff(x\bar b)} {\Bff(xb) + \Bff(x\bar b)} \le  \frac {w(x)} {\Bwf(x)}.  
  \]
  Since these are precisely the terms we care about in the winning condition, Bob wins in the flow-flow game.  
\end{proof}

\begin{lemma}\label{lem:weightflow2flowflow_Alice}
  If Alice has a winning strategy in the weight-flow game, 
  then Alice has a winning strategy in the flow-flow game with the same parameters and for the tree consisting of all strings with zeros in even positions. 
\end{lemma}

\newcommand{\xexp}{x_{\text{exp}}}
\begin{proof}
  Let $T$ be the tree with zeros in even positions, thus $T = \{z : 0 = z_2 = z_4 = \ldots = z_{2\lfloor |z|/2 \rfloor}\}$.\footnote{
    Any tree could be used in which each infinite branch infinitely often alternates between having 1 and 2 children.
    For example, the tree with repeated positions. 
    The above tree was chosen for notational convenience. 
  }
  The idea is to associate strings $x$ in the weight-flow game to expanded strings
  \[
    \xexp = 00x_10x_20 \cdots 0x_{|x|}
  \]
  in the flow-flow game. Alice's weights $w$ in $x$ become flows in $w(\xexp 1)$, since $\xexp 1 \in \partial T$.  
  We explain in detail that this works. 

  Consider a strategy $\Bff$ for Bob in the flow-flow game with the tree from the lemma. 
  Let Bob play the strategy
  \[
    \Bwf(x) = \Bff(\xexp0) + \Bff(\xexp1). 
  \]
  against Alice's winning strategy $w$ in the weight-flow game. 

  Note that the set of $\{\xexp1: x \in \{0,1\}^*\}$ is prefix-free. 
  Alice's strategy in the flow-flow game is obtained by setting
  \begin{align*}
    \Aff(\xexp1)  &= w(x) \text{ for all $x \in \{0,1\}^*$}, \\
    \Aff(\xexp1z) & = 0 \text{ for all $x \in \{0,1\}^*$ and z-nonempty},
  \end{align*}
  and all other values of $\Aff$ are obtained by recursively summing over the values of the children. 
  This implies that $\Aff(\text{empty string}) \le 1$ by the requirement of~$w$. 

  In the winning condition in the flow-flow game, we only sum over prefixes $x$ of odd length, 
  since precisely the nodes $x$ with odd $|x|$ have 1 child. Moreover, this child is always $x0$, 
  thus the winning condition in the flow-flow game only consists of terms of the form
  \[
    \frac {\Aff(\xexp1)} {\Bff(\xexp0) + \Bff(\xexp1)} = \frac {w(x)} {\Bwf(x)}.  
  \]
  Since Alice plays a winning strategy $w$, the sum of $\gamma$-powers of the right-hand side exceeds $s$ and the $\Aff$ strategy is also winning $\Bff$. 
  Because we started with an arbitrary strategy~$\Bff$, the lemma is proven. 
\end{proof}

\section{Bob's strategy in the weight-flow game}

\begin{lemma}\label{lem:upper_bound_eps}
  For each $\eps > 0$, Bob has a winning strategy in the weight-flow game for $\gamma = 1+\eps$ and $s = 2^{\gamma} (\tfrac {8} {(1-2^{-\eps})^3} + 3)$. 
\end{lemma}

\noindent
By lemmas~\ref{lem:weightflow2flowflow_Bob} and~\ref{lem:game2thm_1tree}, this implies the first part of theorem~\ref{th:main_exponential}.

The proof of the lemma starts with an easier variant of the weight-flow game. 
For strings $x,y$, 
let $x \sqsubseteq y$ denote that $x$ is a prefix of~$y$. 
Let $x \sqsubset y$ mean $x \sqsubseteq y$ and $x \not= y$. 

\medskip
\noindent
{\em Variant of the weight-flow game.} The game has an integer parameter~$C \ge 1$. 
Alice and Bob's moves are the same, (Alice increases weights $w$ and Bob a semimeasure~$B$). 
Alice wins if after Bob's move there exist $C$ different strings $x_1, x_2, \ldots, x_C$ such that 
$x_1 \sqsubset x_2 \sqsubset \cdots \sqsubset x_C$ and for all $j = 1, \ldots, C$,
\[
  \frac {w(x_j)} {B(x_j)} > \frac 1 C. 
\]

\begin{lemma}\label{lem:weightflow_easy}
 For each integer $C \ge 1$, Bob has a winning strategy in this variant of the weight-flow game. 
\end{lemma}

\begin{proof}
  With  {\em pushing flow $\delta$ to node $x$}, we mean increasing $B(z) \leftarrow B(z) + \delta$ for all prefixes~$z \sqsubseteq x$. 
  The idea of Bob's reply is to push minimal flows to prevent Alice's winning condition from happening.  
  Thus each time he finds $x_1, \ldots, x_C$ that satisfies this condition, 
  and pushes the minimal flow to $x_C$ so that $w(x_j)/B(x_j) = 1/C$ for some~$j$. 
  In case there are several such $C$-tuples, it does not matter which one is selected first. 
  \begin{algorithm}[H]
    \caption{Bob's reply in the variant of the game.} 
    \begin{algorithmic}[1]
	  \While{strings $x_1 \sqsubset \cdots \sqsubset x_C$ exist with $w(x_j) >B(x_j)/C$ for all $j = 1, \ldots, C$} 
	     \State $\delta\;\;\;\;\;\; \gets \min \{Cw(x_j) - B(x_j): j = 1, \ldots, C\}$
	     \State $B(z) \gets B(z) + \delta$ for all $z \sqsubseteq x_C$ 
	  \EndWhile
    \end{algorithmic}
  \end{algorithm}

  By construction, Alice's winning requirement is never satisfied. 
  We verify that $B$ is always a semimeasure.
  Note that $B(z) \ge B(z0) + B(z1)$ holds for all $z$, because each time a flow $\delta$ is pushed through $z0$ or $z1$, 
  also $B(z)$ is increased. 

  Let $\emptystr$ denote the empty string. 
  It remains to prove that $B(\emptystr) \le 1$. 
  Let $\delta_t$ be the value of $\delta$ in line 2 during the $t$-th push of flow in $B$, (in all replies combined). 
  Now it is explained that
  \[
    \sum_x w(x) \ge \sum_x \max \Big(0,  w(x) - \frac {B(x)}{C} \Big) + \sum_{t} \delta_t. 
  \]
  Initially both sides are zero. 
  At each increase of $w(x)$, the left side can only increase more. 
  Each time some flow $\delta_t$ is pushed, the right side can only decrease, 
  because there are $C$ different strings~$x_j$ where $B(x_j)$ 
  increases by~$\delta_t$ and where~$Cw(x_j) \ge B(x_j) + \delta_t$. 
  Hence, the above inequality always holds. 

  By definition of the game, the left side is at most~$1$, thus $1 \ge \sum_t \delta_t = B(\emptystr)$. 
  Hence, $B$ is a semimeasure and the lemma is proven.  
\end{proof}

\begin{proof}[Proof of lemma~\ref{lem:upper_bound_eps}.] 
  Bob's strategy does not depend on~$\eps$. 
  The idea is that Bob plays the above strategy simultaneously for infinitely many~$C$. 
  To guarantee $B(\epsilon) \le 1$, we need to push slightly smaller flows.  Let \[C_i = i(i+1)2^i.\]
  \begin{algorithm}[H]
    \caption{Bob's reply.} 
    \algtext*{EndFor}{}
    \algtext*{EndWhile}{}
    \algtext*{EndIf}{}
    \begin{algorithmic}[1]
      \While{integer $i\ge 1$ and $x_1 \sqsubset \cdots \sqsubset x_{C_i}$ exist with $w(x_j) >2^{-i}B(x_j)$ for all $j = 1, \ldots, C_i$} 
	     \State $\delta\;\;\;\;\;\; \gets \min \{2^i w(x_j) - B(x_j): j = 1, \ldots, C_i\}$
	     \State $B(z) \gets B(z) + \delta$ for all $z \sqsubseteq x_{C_i}$ 
	  \EndWhile
    \end{algorithmic}
  \end{algorithm}
  As before, $B(z) \ge B(z0) + B(z1)$. To see that $B(\emptystr) \le 1$, consider the sets of rounds $R_i$ where flow is pushed for value~$i$. 
  Let $B = \sum B'_i$ where $B'_i$ corresponds to all flows pushed during rounds in $R_i$. 
  For each $B'_i$, we have 
  \[
    \sum_x w(x) \ge \sum_x \Big| w(x) - \frac {B(x)}{2^i} \Big|_+ + \sum_{t \in R_i}  i(i+1)\delta_t 
  \]
  for exactly the same reasons as in lemma~\ref{lem:weightflow_easy}, 
  but now $B$ gets a $\delta_t$ increase in $C_i = i(i+1)2^i$ places (instead of $2^i$), 
  which implies that the sum decreases by $i(i+1)\delta_t$. 
  Thus, $\sum \{\delta_t : t \in R_i\} \le \tfrac 1 {i(i+1)}$. 
  After summing over $i$, we see that $\sum_t \delta_t \le 1$ and hence, $B$ is indeed a semimeasure. 

  It remains to verify the winning condition. 
  Fix $\omega$. First assume that all prefixes of $x$ satisfy $w(x)/B(x) \le 1$.  
  For each $i$, there are at most $C_i-1$ prefixes~$x$ with $2^{-i} < w(x)/B(x) \le 2^{-i+1}$. 
  Thus,
  \[
    \sum_{x \sqsubset \omega}\left( \frac {w(x)} {B(x)} \right)^{1+\eps}  
    \le \sum_{i=1}^\infty (C_i-1) \cdot (\frac 2 {2^{i}})^{1+\eps} 
    < 2^{1+\eps} \sum_{i=1}^\infty i(i+1)2^{-\eps i} 
    \le \frac {4 \cdot 2} {(1-2^{-\eps})^3}. 
  \]
  Unfortunately, it may happen that $w(x)/B(x) > 1$, but there are at most $C_1 - 1 = 3$ such strings. 
  For this we modify the $B$ flow so that the ratio is always bounded by $2$. Let
  \[
    B'(x) = \tfrac 1 2 B(x) + \tfrac 1 2 \sum_{z:x \sqsubseteq z} w(z). 
    \]
    The ratio $w(x)/B'(x) \le 2$. The sum for $B'$ contains an additional factor $2^\gamma$ plus the contribution of at most $3$ larger terms. 
    Hence, Bob wins the game for 
    \[
      s = 2^\gamma (\frac {8} {(1-2^{-\eps})^3} + 3).
    \]
  Lemma~\ref{lem:upper_bound_eps} is proven.  
\end{proof}

\begin{lemma}\label{lem:upper_bound_gamma1}
  Bob has a winning strategy in the weight-flow game for $\gamma = 1$ and some $s = O(\log^2 N \log^4 \log N)$. 
\end{lemma}

\begin{proof}
  We start by analysing the finite game restricted to $N$-bit strings where $N \ge 2$. 
  The algorithm is the same as for lemma~\ref{lem:upper_bound_eps}, but using better coefficients $C_i = \lceil \log N\rceil 2^i$. 

  For notational convenience, we ignore the first few terms in the sum of the winning condition, 
  equivalently, we assume $0 = w(\epsilon) = w(0) = w(1) = \ldots$, since finitely many terms do not matter given the big-O notation in~$s$.  
  For this, we need that the ratios $w(x)/B(x)$ are bounded on those finitely many strings, 
  and this is achieved by adding a uniform flow to Bob's flow and rescaling proportionally. 
  Note that this rescaling factor is absorbed in the big-$O$.

  Note that only flows are pushed for values of $i$ with $C_i \le N$, thus $i \le \lceil \log N \rceil$. 
  If all prefixes of $x$ satisfy $w(x)/B(x) \le 1$, then
  \[
    \sum_{x \sqsubset \omega_{\le N}} \frac {w(x)} {B(x)}
    \le \sum_{i=1}^{\lceil \log N\rceil} (C_i-1) \cdot \frac 2 {2^{i}}
    < O(\log^2 N). 
  \]
  There might be $C_1 \le O(\log N)$ prefixes $z$ for which $w(z)/B(z) > 1$, but in a similar way as before, we bound their ratio by~$2$. 

  Now we need a strategy for all lengths~$N$. 
  For this, we run infinitely many strategies of Bob in parallel where each strategy $w_\ell$ is supported 
  on strings whose lengths are in the interval $[M_\ell, M_{\ell + 1}-1]$ where
  \[
    M_\ell = 2^{2^\ell}.
  \]
  Thus, $w_\ell(x) = w(x)$ if $M_\ell \le |x| < M_{\ell + 1}$ and 0 otherwise. 
  Bob's strategy strategy is 
  \[
    B(x) = \sum_\ell \frac {B_\ell(x)} {\ell(\ell + 1)}. 
  \]
  We verify that Bob is winning.
  \begin{align*}
    \sum_{x \sqsubset \omega_{\scaleobj{0.8}{\le}N}} \frac {w(x)} {B(x)}
    & \le \sum_{\ell=1}^{\lceil\log \log N\rceil} \sum_{n = M_\ell}^{M_{\ell+1}-1} \ell(\ell+1) \frac {w_\ell(\omgle)} {B_\ell(\omgle)} \\
    & \le \sum_{\ell=1}^{\lceil\log \log N\rceil} \ell^2 \cdot O(\log^2 M_\ell). 
  \end{align*}
  Finally, we observe that $\log M_\ell$ increases geometrically, thus
  \[
    < O(\log^2 N \log^2 \log N). 
    \qedhere
  \]
\end{proof}

\begin{proof}[Proof of theorem~\ref{th:main_exponential}.] 
  Apply lemmas~\ref{lem:weightflow2flowflow_Bob}, \ref{lem:game2thm_1tree} and~\ref{lem:upper_bound_eps} 
  (respectively lemma~\ref{lem:upper_bound_gamma1}) to obtain the proof of the first part (respectively second part). 
  Let $p$ be a program that decides the tree $T$. Then, we have that $\K(p) \le |p| + 2\log |p|$. 
  Thus $2^{\K(T)} \le k^2 2^k$. The theorem is proven. 
\end{proof}

\section{Upper bound for online classification}

Let $\langle z \rangle$ be a computable prefix-free encoding of string $z$. 
Recall that we view the task of learning $f : \{0,1\}^* \rightarrow \{0,1\}$ through online classification as a sequence prediction task. 
The learner iteratively receives $x_n$, predicts $y_n$ and receives the correct value $f(x_n)$. 
This task is described as a sequence prediction task of selected bits on a sequence
\[
\langle x_1 \rangle f(x_1) \langle x_2 \rangle f(x_2) \cdots
\]
Let $T_f$ be the leafless tree consisting of all prefixes of such sequences for all choices of $x_1, x_2, \ldots \in \{0,1\}^*$. 

\begin{lemma}\label{lem:BobWinsClassificationGame}
  Fix $\gamma > 1$. 
  If $T_1, \ldots, T_{2^k}$ are trees obtained in the above way from classifiers, 
  then Bob has a computable winning strategy for some $s \le O(k^{1+\gamma})$. 
\end{lemma}

\begin{proof}[Proof of theorem~\ref{th:classifiers_upper} assuming lemma~\ref{lem:BobWinsClassificationGame}.] 
  Let $T_1, \ldots, T_{2^k}$ be a list of leafless trees that contain all $T_f$ for classifiers $f$ 
  that are computed by programs of length less than~$k$ within time bound~$t$. 
  Such a list can be made uniformly computable in $k$
  by applying the same technique as in the the construction of maximal semimeasures: 
  interpret each string of length less than $k$ as a program that 
  computes such a tree, possibly by fixing the output if a program terminates too slowly or gives invalid output (for example, produces a leaf). 
  Theorem~\ref{th:classifiers_upper} follows from lemma~\ref{lem:game2thm} applied to trees $T_1, \ldots, T_{2^k}$ 
  and the strategy from lemma~\ref{lem:BobWinsClassificationGame}. 
\end{proof}

\medskip
\noindent
In the remainder of this section, lemma~\ref{lem:BobWinsClassificationGame} is proven.  
Recall the transition from the flow-flow game with 1 tree $T$ to the weight-flow game in the previous section:
when Alice's flow leaves the tree $T$, this corresponds to increasing the weight on the node from which the flow leaves. 

A naive way to transit from a flow-flow game with many trees to the weight-flow game is to is to increase 
the weights of a node in which flow leaves some tree. 
To guarantee that the sum of weights is at most $1$, the weights need to be rescaled by the number of trees. 
This means that the strategy of the weight-flow game requires $s$ proportional to~$2^k$. 
This is not enough, because we need~$O(k^{1+\gamma})$. 
The idea is to partition the trees into a small number of forests, and only increase the weights when the flow leaves 1 of the forests. 

\begin{definition}\label{def:}
  A  {\em leafless forest} $F$ is a subset of $\{0,1\}^*$ such that for all $x \in F$, either $x0 \in F$ or $x1 \in F$. 
\end{definition}

\noindent
Note that a tree obtained from a classifier, satisfies this property. 

\begin{definition}\label{def:}
  A tuple $F_1, \ldots, F_k$ of forests is called a {\em good partition} of the leafless trees $T_1, \ldots, T_K$~if
  \\-- $F_1, \ldots, F_k$ partitions all nodes in $T_1 \cup \ldots \cup T_K$, 
  \\-- forests $F_j$ are leafless,
  \\-- if a node $v$ in a tree $T_i$ has a single child, then $v$ has a single child in the forest $F_j$ containing $v$,
  \\-- the intersection of a branch $\omega$ and a forest $F_i$ is either empty or a path, 
  thus $\{n : \omgle \in F_i\}$ is an interval in $\mathbb N$ (possibly empty or infinite). 
\end{definition}

\begin{lemma}\label{lem:flowflow2weightflow_Bob_manyTrees}
  Assume forests $F_1, \ldots, F_k$ are a good partition of leafless trees $T_1, \ldots, T_K$. 
  Also, assume that Bob has a winning strategy in the weight-flow game with parameters $\gamma$ and~$s$. 
  Then Bob has a winning strategy in the flow-flow game on the trees $T_1, \ldots, T_K$ with parameters $\gamma$ and $2^\gamma k(1 + k^\gamma s)$. 
\end{lemma}

\noindent
First this lemma is proven, and afterwards it is explained how forests $F_i$ are obtained. 

\begin{figure}[tb]
  \centering
  \begin{tikzpicture}[
    x=1cm,y=1cm,
    font=\small,
    line cap=round,line join=round,
    fone/.style={draw=black!65,line width=.8pt},
    ftwo/.style={draw=lemgreen!65,line width=.8pt},
    crossing/.style={draw=black!45,line width=.8pt,
		     dash pattern=on 3pt off 2.5pt},
    branch/.style={draw=lemgreen,line width=1.8pt},
    dot/.style={circle,inner sep=0pt,minimum size=3.8pt,
		fill=black!75,draw=white,line width=.25pt},
    fdot/.style={dot,fill=lemgreen!70},
    marked/.style={circle,inner sep=0pt,minimum size=7pt,
		   draw=white,line width=.6pt},
    rootmark/.style={marked,fill=lemgreen},
    callout/.style={rectangle,rounded corners=3pt,fill=white,
		    line width=.75pt,inner xsep=5pt,inner ysep=4pt,
		    align=center},
    continuation/.style={dash pattern=on 1.4pt off 2.4pt}
  ]
    \definecolor{lemblue}{RGB}{38,96,171}
    \definecolor{lemred}{RGB}{185,47,51}
    \definecolor{lemgreen}{RGB}{20,94,61}

    \coordinate (r)   at ( 0.00, 0.00);
    \coordinate (a)   at (-0.25,-0.75);
    \coordinate (b)   at ( 0.00,-1.50);
    \coordinate (c)   at (-2.05,-2.25);
    \coordinate (x)   at (-2.30,-3.00); 
    \coordinate (y)   at (-3.20,-3.75); 
    \coordinate (l1)  at (-3.35,-4.50);
    \coordinate (l2)  at (-3.10,-5.25);
    \coordinate (ll1) at (-3.85,-6.00);
    \coordinate (ll2) at (-3.95,-6.75);
    \coordinate (lr1) at (-2.40,-6.00);
    \coordinate (lr2) at (-2.35,-6.75);
    \coordinate (d)   at ( 2.00,-2.25);
    \coordinate (e)   at ( 1.70,-3.00);
    \coordinate (f)   at ( 1.90,-3.75);
    \coordinate (h1)  at ( 1.15,-4.50);
    \coordinate (h2)  at ( 1.00,-5.25);
    \coordinate (h3)  at ( 1.15,-6.00);
    \coordinate (h4)  at ( 1.10,-6.75);
    \coordinate (j1)  at ( 2.65,-4.50);
    \coordinate (j2)  at ( 2.50,-5.25);
    \coordinate (j3)  at ( 2.70,-6.00);
    \coordinate (j4)  at ( 2.60,-6.75);

    \coordinate (z)   at (-1.10,-3.75);
    \coordinate (p1)  at (-0.75,-4.50);
    \coordinate (p2)  at (-0.90,-5.25);
    \coordinate (p3)  at (-1.45,-6.00);
    \coordinate (p4)  at (-1.40,-6.75);
    \coordinate (s1)  at (-0.15,-6.00);
    \coordinate (s2)  at (-0.05,-6.75);

    \coordinate (u)   at ( 3.45,-3.00);
    \coordinate (u1)  at ( 3.90,-3.75);
    \coordinate (u2)  at ( 4.10,-4.50);
    \coordinate (v1)  at ( 3.75,-5.25);
    \coordinate (v2)  at ( 3.65,-6.00);
    \coordinate (v3)  at ( 3.80,-6.75);
    \coordinate (w1)  at ( 4.90,-5.25);
    \coordinate (w2)  at ( 5.05,-6.00);
    \coordinate (w3)  at ( 4.95,-6.75);

    \path[fill=lemgreen!5,rounded corners=7pt]
      (-1.36,-3.53) -- (-0.84,-3.53) -- (-0.45,-4.46)
      -- (-0.58,-5.15) -- (0.18,-5.85) -- (0.31,-7.31)
      -- (-1.78,-7.31) -- (-1.78,-5.91) -- (-1.19,-5.12)
      -- (-1.06,-4.53) -- cycle;
    \path[fill=lemgreen!5,rounded corners=7pt]
      (3.19,-2.78) -- (3.67,-2.78) -- (4.18,-3.62)
      -- (4.40,-4.38) -- (5.18,-5.08) -- (5.37,-7.31)
      -- (3.44,-7.31) -- (3.30,-5.89) -- (3.43,-5.13)
      -- (3.80,-4.45) -- (3.61,-3.88) -- cycle;

    \foreach \v/\w in {
      r/a,a/b,b/c,c/x,x/y,y/l1,l1/l2,l2/ll1,ll1/ll2,
      l2/lr1,lr1/lr2,b/d,d/e,e/f,f/h1,h1/h2,h2/h3,h3/h4,
      f/j1,j1/j2,j2/j3,j3/j4}
      \draw[fone] (\v)--(\w);
    \foreach \v/\w in {
      z/p1,p1/p2,p2/p3,p3/p4,p2/s1,s1/s2,
      u/u1,u1/u2,u2/v1,v1/v2,v2/v3,u2/w1,w1/w2,w2/w3}
      \draw[ftwo] (\v)--(\w);

    \draw[crossing] (d)--(u);

    \draw[branch] (r)--(a)--(b)--(c)--(x);
    \draw[branch,dash pattern=on 3pt off 2.5pt] (x)--(z);
    \draw[branch] (z)--(p1)--(p2)--(p3)--(p4);
    \draw[branch,continuation,-{Stealth[length=4pt,width=4pt]}]
      (p4)--(-1.40,-7.60);
    \node[text=lemgreen,anchor=west,font=\normalsize]
      at (-1.18,-7.49) {$\omega$};

    \foreach \v in {ll2,lr2,h4,j4}
      \draw[fone,continuation] (\v)--++(0,-.56);
    \foreach \v in {s2,v3,w3}
      \draw[ftwo,continuation] (\v)--++(0,-.56);

    \foreach \v in {r,a,b,c,l1,l2,ll1,ll2,lr1,lr2,d,e,f,
		      h1,h2,h3,h4,j1,j2,j3,j4}
      \node[dot] at (\v) {};
    \foreach \v in {p1,p2,p3,p4,s1,s2,u1,u2,v1,v2,v3,w1,w2,w3}
      \node[fdot] at (\v) {};
    \node[marked,fill=lemblue] at (x) {};
    \node[marked,fill=lemred] at (y) {};
    \foreach \v in {z,u} {
      \draw[lemgreen,line width=.65pt] (\v) circle[radius=4.6pt];
      \node[rootmark] at (\v) {};
    }

    \node[anchor=south,inner sep=4pt] at (r) {$\epsilon$};
    \node[font=\large,anchor=west] at (0.28,-0.71) {$F_1$};
    \node[text=lemgreen,font=\large,anchor=east]
      at (-1.58,-4.94) {$F_2$};
    \node[text=lemgreen,font=\large,anchor=west]
      at (4.47,-4.05) {$F_2$};
    \node[text=lemgreen,anchor=south west,inner sep=3pt]
      at (3.62,-2.98) {another root of $F_2$};

    \node[callout,draw=lemblue,text=lemblue,anchor=east]
      (bluebox) at (-2.92,-2.96) {$\omega_{<n}\in F_1$};
    \draw[lemblue,line width=.65pt] (bluebox.east)--(x);
    \node[callout,draw=lemred,text=lemred,anchor=east]
      (redbox) at (-3.79,-3.90)
      {$\omega_{<n}\bar\omega_n\in F_1$\\
       $\omega_{<n}\bar\omega_n\not\in T_i$};
    \draw[lemred,line width=.65pt] (redbox.east)--(y);

    \node[callout,draw=lemgreen,text=lemgreen,anchor=west]
      (greenbox) at (-0.51,-3.64)
      {$\omega_{\le n}\in F_2$\\[-1pt]
       {\footnotesize root of $F_2$}};
    \draw[lemgreen,line width=.65pt] (z)--(greenbox.west);
    \node[text=lemred,anchor=west,inner sep=1pt]
      at (-2.78,-3.60) {$\bar\omega_n$};
    \node[text=lemgreen,anchor=west,inner sep=1pt]
      at (-1.64,-3.27) {$\omega_n$};
  \end{tikzpicture}
  \caption{%
    The branch $\omega$ leaves $F_1$ through a root of $F_2$.
    The red child $\omega_{<n}\bar\omega_n$ belongs to $F_1$
    but not to $T_i$, so this index contributes to the
    $T_i$-sum but not to the $F_1$-sum.%
  }
  \label{fig:winningcondition_forests}
\end{figure}
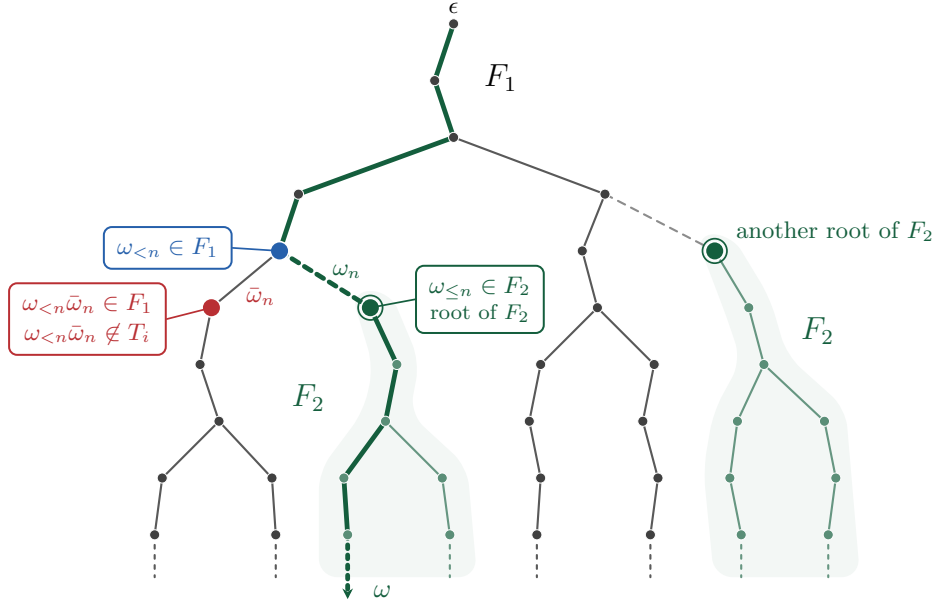

\begin{proof}
  We will use $k$ copies of the weight-flow, where
  for Alice, a strategy $w_j$ in the weight-flow game is obtained by increasing~$w$ 
  for each flow that leaves forest $F_j$ from a node in~$F_j$. 
  For Bob, a strategy is obtained in the flow-flow game by pushing the flow 1 step forward on nodes with 1 child in its forest. 
  Here are the details. 

  Let $\Aff$ be a move by Alice in the flow-flow game. Recall that $F_j$ is leafless.  

  \medskip
  \noindent
  {\em Alice's move $w_j$ in the $j$-th weight-flow game.} 
  Let $w_j(x) = \Aff(xb)$ if $x \in F_j$ and $xb \in \partial F_j$ for some bit~$b$. For all other $x$, let $w_j(x) = 0$. 

  \medskip
  \noindent
  Note that $\sum_x w_j(x) \le 1$, because $\partial F_i$ is prefix-free. 
  
  \medskip
  \noindent
  {\em Bob's move in the flow-flow game.} 
  Let $\Bwfj$ be the reply from Bob when he plays a winning strategy against Alice's move~$w_j$. 
  Flow $\Bff$ is obtained by pushing each flow $\Bwfj$ for 1 step forward on nodes $p$ that have a unique child $pb$ in the direction of $pb$, 
  and taking the average over these $\Bwfj$ and $\Aff$'s flows. 

  Formally, let $\Bffj'(x) = \Bwfj(p)$ if $p$ is the parent of $x$ and $p$ has exactly 1 child $x$ in its forest. If $p$ has 2 children, then $\Bffj'(x) = \Bwfj(x)$. 
  If $x$ is on a path to at least 1 root of forest $F_j$, then let $\Bffj'(x)$ be the sum of all flows $\Bwfj(r)$ over roots~$r$ that are extensions of~$x$. 
  If $x$ is neither on $F_j$ nor on a path to some of its roots, then $\Bffj'(x) = 0$. 
  Finally, for all $x$, we set 
  \[
    \Bff(x) = \tfrac 1 2\Aff(x) + \tfrac 1 {2k} \sum_{j \le k} \Bffj'(x). 
  \]

  \medskip
  \noindent
  For the same reasons as before, $\Bffj'$ is a semimeasure, thus so is~$\Bff$. 
  It remains to verify that Bob wins the game with parameters $\gamma$ and $k2^\gamma (1 + k^\gamma s)$. 

  Fix a tree $T_i$ and an infinite branch $\omega$. We need to bound
  \[
    \sum_{\substack{n: \omglt \in T_i \\ \omglt \bar\omega_n \not\in T_i}} 
    \left( \frac {\Aff(\omglt \bar\omega_n)} {\Bff(\omglt \omega_n) + \Bff(\omglt \bar \omega_n)} \right)^\gamma.
  \]
  We rewrite this so that the winning condition of the weight-flow game can be used, which requires to think about forests.  

  All flow that leaves tree $T_i$ also leaves some forest $F_j$, except in those nodes where the tree $T_i$ leaves a forest $F_j$,
  see figure~\ref{fig:winningcondition_forests}.  
  This exceptional flow stays in the forest, and this flow is not accounted for in any game. There is 1 such term per forest. 
  \[
    \sum_{\substack{n: \omglt \in T_i \\ \omglt \bar\omega_n \not\in T_i}} \cdots \quad 
    \le \quad \sum_{j \le k} \textnormal{Missing}_j  + \sum_{\substack{n: \omglt \in F_j \\ \omglt \bar\omega_n \not\in F_j}} \cdots 
  \]
  Since $\Bff \ge \tfrac 1 2 \Aff$, each (missing) term is at most $2^\gamma$. Hence, we need to bound
  \[
    \le 2^\gamma k  + \sum_{j \le k} \sum_{\substack{n: \omglt \in F_j \\ \omglt \bar\omega_n \not\in F_j}} 
    \left( \frac {\Aff(\omglt \bar\omega_n)} {\Bff(\omglt \omega_n) + \Bff(\omglt \bar \omega_n)} \right)^\gamma.
  \]
  Note that if $x$ has 1 child $xb$ in $F_j$, then $\Bwfj(x) = \Bffj'(xb)$. 
  By definition of $w_j$, 
  \[
    \le 2^\gamma k  + \sum_{j \le k} \sum_{\substack{n: \omglt \in F_j \\ \omglt \bar\omega_n \not\in F_j}} 
    \left( \frac {w_j(\omglt)} {\tfrac 1 {2k} \Bwfj(\omglt)} \right)^\gamma.
  \]
  Now the lemma finishes by using that Bob is winning the weight-flow games, thus
  \[
    \le 2^\gamma k + k \cdot (2k)^\gamma s. 
    \qedhere
  \]
\end{proof}

\bigskip
\noindent
It remains to construct the forests. 

\begin{definition}\label{def:}
  Let $T$ be a tree. 
  \\-- A node $z \in T$ is called {\em splitting} if both $z0 \in T$ and $z1 \in T$. 
  \\-- A node $z \in T$ is called {\em predicting} if it has exactly 1 child in~$T$.  
  \\-- Trees $T_1, \ldots, T_K$ are called {\em jointly predicting} if for every $z$, either 
  all trees that contain $z$ are splitting or all are predicting. 
\end{definition}

\noindent
Observe that a set of trees $T_f$ associated to classifiers is jointly predicting. 
Indeed, if 2 trees $T_f$ and $T_g$ contain a node $z$, 
it means that both $f$ and $g$ are consistent with the labels in $z$. 
When extending $z$, this remains true as long as $f$ and $g$ agree on the encoded inputs. 
If $z = \langle x_1 \rangle f(x_1) \ldots \langle x_{n-1} \rangle f(x_{n-1}) \langle x_{n} \rangle$ for some 
$x_n$ with $f(x_n) \not= g(x_n)$, then both trees have exactly $1$ child, but it will be a different child.

\begin{lemma}\label{lem:getForests}
  Given $2^k-1$ leafless trees $T_1, \ldots, T_{2^k-1}$ that are jointly predicting, 
  one can compute $k$ forests $F_1, \ldots, F_k$ that are a good partition of these trees. 
\end{lemma}

\noindent
The idea of the construction 
mimics the halving algorithm for online classification given in~\cite{Littlestone1988learning}.

First we construct the `most popular' subtree $F_1$ in a greedy way, starting at the root: 
if the current node is splitting, add it to $F_1$ and recurse on both children, otherwise, 
recurse on the child that lays on the maximal number of trees $T_j$ (ignore the other child). 
Break ties arbitrarily. 
Then, remove $F_1$ from $U$. Now $U$ is a forest. 
We repeat the process for each root in $U$ to obtain forest $F_2$ (thus $F_2$ consists of the ``most popular subtrees'' in the remaining~$U$). 
Then remove $F_2$ from $U$ and repeat to construct $F_3$, and so on. 
Note that this is a mathematical construction, not an algorithm. 

\begin{algorithm}[H]
  \captionsetup{labelformat=empty}
  \caption{Construction of forests $F_1, \ldots, F_k$ for fixed $T_1, \ldots, T_K$}
  \algtext*{EndFor}{}
  \algtext*{EndWhile}{}
  \algtext*{EndIf}{}
  \textsc{ForestConstruction()}
  \begin{algorithmic}[1]
    \State $U \gets T_1 \cup \cdots \cup T_K$. 
    \For{$i \gets 1,2,\ldots,k$}
      \State $F_i \gets \emptyset$
      \For{every \textit{root} in $U$}
	  \State \textsc{Process}(\textit{root}, $U$, $F_i$)
      \EndFor
      \State remove $F_i$ from $U$
  \EndFor
  \State Output $F_1, \ldots, F_k$.
  \end{algorithmic}

  \medskip
  \textsc{Process}(\textit{node}, $U$, $F$)
  \begin{algorithmic}[1]
    \While{\textbf{true}}
	\State add \textit{node} to $F$ 
	\If{\textit{node} is a splitting node}
	   \State \textsc{Process}(\textit{child 1}, $U$, $F$)
	   \State \textsc{Process}(\textit{child 2}, $U$, $F$)
	\Else  $\;\;$ \color{gray} \# thus node has 1 child in each $T$ \color{black}
	   \State \textit{child} $\gets$ child on the maximal number of trees $T_j$
	   \State \textsc{Process}(\textit{child}, $U$, $F$)
	\EndIf
    \EndWhile
  \end{algorithmic}
\end{algorithm}

\begin{lemma}
  The output $F_1, \ldots, F_k$ of \textsc{ForestConstruction} partitions $T_1 \cup \cdots \cup T_K$. 
\end{lemma}

\begin{proof}
  \textbf{Claim.} 
  {At the beginning of the $i$-th iteration of the for loop in \textsc{ForestConstruction}, 
  every remaining node in $U$ lies on less than $2^{k + 1 - i}$ trees.} 

  \medskip
  \noindent
  Proof of the claim.
  Since the root lays on less than $2^k$ trees, this holds for $i = 1$. 

  Now inductively assume the claim for some $i$, thus we assume every node in $F_i$ 
  lays on less than $2^{k+1-i}$ trees. 
  Every node in $F_{i+1}$ was not added to $F_i$, and such a decision can happen only in the else case inside~\textsc{Process}. 
  It happens only for the child that lays on the smaller number of nodes. Thus this number is smaller than $2^{k+1-i}/2 = 2^{k+1-i-1}$. 
  The inductive step and hence the claim are proven. 

  \medskip
  \noindent
  This implies that at the beginning of the $k$-th iteration, every node in $U$ lies on exactly 1 tree and \textsc{Process} copies $U$ into $F_k$. 
  Hence, the forests $F_1, \ldots, F_k$ cover all the trees. 
\end{proof}

\begin{proof}[Proof of lemma~\ref{lem:getForests}.] 
  It is easy to see that the forests are computable, because the decision to include a node $x$ in some forest only depends on 
  the part of the trees on strings of length at most~$|x|$. 

  It remains to verify that $F_1, \ldots, F_k$ is a good partition of $T_1, \ldots, T_K$. 
  The previous lemma states that it is a partition. 

  At any moment, $U$ is leafless because the copying to $F_j$ always starts from a root in~$U$. 
  Hence, each $F_{j+1}$ is leafless, because the most popular child always exists and the copying never terminates. 

  If a node $v$ contains 1 child in a tree $T_i$, then this node is predicting and has 1 child in the forest containing~$v$. 

  Finally, observe that each infinite branch $\omega$ of a tree $T_i$ intersects $F_j$ in a path. 
  Indeed, each ``less popular child'' finishes the copying of the prefix to the current~$F_j$. 
  Afterwards, no more prefixes are copied from~$\omega$. 

  In conclusion, the forests are a good partition. 
  Lemma~\ref{lem:getForests} and hence theorem~\ref{th:classifiers_upper} are proven.  
\end{proof}

%
%
%
%

\section{Alice's strategy in the weight-flow game}

\noindent
Note that we consider essentially finite games now. 

\begin{lemma}\label{lem:lowerbound_strategy}
  For each~$N$, Alice has a strategy in the weight-flow game in which for some $N$-bit string $z$ she achieves the following. 
  For infinitely many rounds,
  \begin{equation}\tag{$*$}\label{eq:winning}
    \sum_{n \le N} \frac {w(z_{\le n})} {B(z_{\le n})} \;>\;\Omega(\log N).
  \end{equation}
  Moreover, these strategies are computable given~$N$. 
\end{lemma}

\begin{proof}[Proof of  theorem~\ref{th:main_L1}.] 
  Select $N$ of the form $2^{2^\ell}$. Note that $2^{\K(N)} \le O(\log^2 \log N)$.  
  The theorem now follows from lemmas~\ref{lem:game_lowerbound} and~\ref{lem:weightflow2flowflow_Alice}. 
\end{proof}

\noindent
Thus it remains to prove the lemma. 

\begin{figure}
  \centering
  \begin{tikzpicture}[
    x=1cm, y=1cm,
    line width=0.9pt,
    line join=round,
    tip/.style={circle, fill=black, inner sep=0pt, minimum size=5pt},
    scale=0.67
  ]
    \coordinate (outerLeft)  at (-4.9,  1.4);
    \coordinate (outerRight) at ( 4.9,  1.4);
    \coordinate (outerTip)   at ( 0,   -6);
    \filldraw[fill=black!30]
      (outerLeft) -- (outerRight) -- (outerTip) -- cycle;

    \coordinate (whiteLeft)  at (-3.7,  0);
    \coordinate (whiteRight) at ( 3.7,  0);
    \coordinate (whiteTip)   at ( 0,   -5.6);
    \filldraw[fill=white]
      (whiteLeft) -- (whiteRight) -- (whiteTip) -- cycle;

    \foreach \x/\lab in {-3.2/1, -1.6/2, 3/n} {
      \fill
	({\x-0.62},1.25) -- ({\x+0.62},1.25) -- (\x,0) -- cycle;
      \node[tip] (tip-\lab) at (\x,0) {};

      \node[anchor=base, inner sep=0pt]
	at (\x,-0.55) {$\lab$};

      \node[anchor=west, inner sep=1pt]
	at ({\x+0.46},0.625) {$s$};
    }

    \node at (0.75, 0.625) {$\cdots$};
    \node at (0.75,-0.37)  {$\cdots$};

    \draw[<->, >=Stealth, line width=0.8pt,
	  shorten <=3pt, shorten >=3pt]
      (0,0) -- (whiteTip)
      node[midway, right=4pt] {$\lceil 5s\rceil$};

    \path[darkgray] (outerRight) -- (outerTip)
      node[midway, right=8pt] {$s+\varepsilon$};
  \end{tikzpicture}
  \caption{concatenation of strategies in the proof of lemma~\ref{lem:lowerbound_strategy}.} 
  \label{fig:concatenate_lowerbound_strategies}
\end{figure}
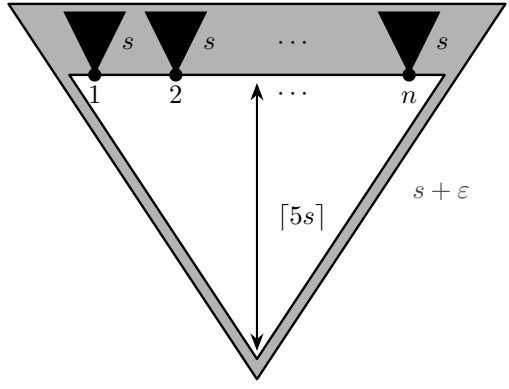

\begin{proof}
  This proof follows the main technical result in~\cite{Andreev2016sum}, which is their lemma~4.  
  We note that their strategy is uniformly computable and has the following property. 
  After every move of Alice and before Bob's move, Alice is in a winning position in a certain string $x$. 
  If Bob invalidates this winning position, then Alice will never attempt to recover the winning position in $x$, but instead create a winning position in another string. 
  The proof shows that eventually, some string becomes winning for Alice and remains winning forever.


  We only explain the changes, which involves estimating the error when approximating a finite sum by an integral. 

  The construction of the strategy is recursive in the target~$s$: given a strategy for parameter $s$, 
  a new strategy is obtained with parameter $s + 2^{-\lceil 4s\rceil}$ using strings of length at most~$\lceil 5s\rceil$ and playing the recursive strategies at the end of each string, 
  see figure~\ref{fig:concatenate_lowerbound_strategies}. 

  Let the gain be $\eps = 2^{-\lceil 4s\rceil}$ and the number of strings be~$n = 2^{\lceil 5s\rceil}$. 
  For this, we need to find values of $\beta_i$ and $d_i$ such that $\sum \beta_i \ge 1$ where 
  both quantities are obtained from $(*)$ and $(**)$ on page 11 of~\cite{Andreev2016sum}. 
  Here $\beta_i$ is the bound of the game on the $i$-th string.
  The equations are 
  \begin{align*}
    \frac{s(1-\eps)(1-\tfrac i n)} {1-d_i} & = s + \eps \\
    s + \eps - \tfrac {\eps} {d_i} = s \frac {(1-\eps)/n} {\beta_i}. 
  \end{align*}
  Let $t_i = \eps + \tfrac i n (1-\eps)$. 
  We have
  \[
    \beta_i = \frac {(1-\eps)(\eps + st_i)} {n(s + \eps)t_i}. 
  \]
  Note that $\beta_i$ is decreasing in~$i$.  One has 
  \[
    \sum_{i=1}^n \beta_i = \frac  {1-\eps} {n(s+\eps)} \sum_{i=1}^n \frac {\eps + st_i} {t_i}  \ge \frac  {1} {s+\eps} \int_\eps^1 \frac {\eps + su}{u} \mathrm d u - \beta_0 + \beta_n.
  \]
  Solving the integral gives
  \[
    = \frac {\eps \ln \tfrac 1 \eps + (1-\eps)s} {s+\eps} - \beta_0 + \beta_n.
  \]
  One verifies that for $\eps = 2^{-\lceil 4s\rceil}$, $n = 2^{\lceil 5s\rceil}$ and $s \ge 1$, this bound is indeed at least~$1$. 

  It remains to verify that the length of the strings needed for this stratagey is $2^{cs}$ when $s \ge c$. 
  Indeed, to increase $s$ from $i-1$ to $i$ (with $i$ integer), 
  we need to repeat this construction at most $2^{4i}$ times. 
  Thus, to reach a value $s-1$, the total length of the concatenated strings is 
  \[
    \sum_{i = 1}^{s} 5i \cdot 2^{4i} \le O(s 2^{4s}). 
  \]
  Thus there exists a constant $c$ such that this is at most $2^{c(s-1)}$ if $s-1 \ge c$. 
   Lemma~\ref{lem:lowerbound_strategy} is proven.  
\end{proof}

\begin{proof}[Proof of corollary~\ref{cor:classifiers_lower}.]
  Recall that in the proof of  lemma~\ref{lem:weightflow2flowflow_Alice}, 
  we can choose a tree in which the predicted bits are precisely those in even position, 
  and the prediction is always zero.
  This tree corresponds to any classifier that maps both strings ``0'' and ``1'' to the bit $0$. 
  This almost works, but there is a small problem, because we need to assume a prefix encoding of the strings in the input stream. 
  But the fix is obvious, the tree should alternate predicting layers with several branching layers according to the prefix encoding. 
\end{proof}

\subsubsection*{Statement about AI usage}

All theorems were proved and written by the authors. 
Theorems~\ref{th:main_exponential} and \ref{th:main_L1} are based on German's thesis in from June 2025. 
Tikz pictures were generated by chatgpt 6.1 before the public release. 
Chatgpt was also asked to write a detailed review. It gave useful references to the halving algorithm and Hutter's work. 
It foudn a major mistake of a deleted theorem, as well as a large number of fixable mistakes.

\bibliographystyle{plain}
\bibliography{kolmogorov}

\appendix

\section{Online semimeasures are slightly better for online classification}
\label{sec:online}

\begin{definition}[\cite{onlineComplexity}\label{def:evenSemimeasure}]
  An {\em online} semimeasure $\Pon$ is a function that maps lists of the form $x_1, b_1, x_2, b_2, \ldots, x_n$ or $x_1, b_1, x_2, b_2, \ldots, x_n, b_n$ 
  with $x_i \in \{0,1\}^*$ and $b_i \in \{0,1\}$ to $\mathbb R_{\ge 0}$ that satisfy
     \begin{flalign*} 
       \;\;\bullet\;\; & \Pon(\textnormal{empty list})  \le 1, & \\ 
       \;\;\bullet\;\; & \Pon(x_1, b_1, \ldots, x_{n-1}, b_{n-1})  = \Pon(x_1, b_1, \ldots, x_{n-1}, b_{n-1}, x_n), & \\ 
       \;\;\bullet\;\; & \Pon(x_1, b_1, \ldots, x_{n})  \ge \sum_{\!\!b_n \in \{0,1\}\!\!} \Pon(x_1, b_1, \ldots, x_n, b_n). & 
  \end{flalign*}
\end{definition}

\noindent
{\em Online} algorithmic probability is defined 
in a similar way as standard algorithmic probability by the following result. 

\begin{proposition}[\cite{onlineComplexity}]\label{prop:algorithmicOnlineProbability}
  Among the lower semicomputable online 
  semimeasures, there exists a semimeasure $\Mon$ that is maximal up to a constant factor, 
  thus for each such semimeasure $\Pon$ one has $\exists c \forall L[\Pon(L) \le c \Mon(L)]$. 
\end{proposition}

\noindent
All $\Mon$ that satisfy the conditions of the proposition are equal up to a constant factor. We fix such~$\Mon$. 
An interesting application is Marcus Hutter's AIXI policy for reinforcement learning~\cite{Hutter2005universal, Hutter2024book}.
This theoretical policy models the environment using $\Mon$ and practical variants of the policy have been studied~\cite{Hutter2024book}.\footnote{
  Hutter uses the notion of chronological measure, see \cite[definition 7.2.1-7.2.2]{Hutter2024book} 
  and sets the weights as in definition 3.7.2. 
  This provides an optimal lower semicomputable online semimeasure when model class $\mathcal M$ consists of all lower semicomputable semimeasures. 
  This construction follows the proof of proposition~\ref{prop:algorithmicOnlineProbability}.  
}

\medskip
\noindent
{\em Side note.} The {\em online} Kolmogorov complexity of $x_1, b_1, \ldots, x_n, b_n$ is the 
minimal length of a program that for each $i \le n$ computes $b_i$ given $x_1, \ldots, x_{i}$. 
An online variant of the coding theorem holds: 
the negative logarithm of online algorithmic probability equals online complexity up to additive $O(\log n)$ terms~\cite[theorem 2]{onlineComplexity}. 
Unfortunately, symmetry of information does not hold~\cite{Bauwens2014asymmetry} 
but does hold for spacebounded complexity with reasonable precision~\cite{Bauwens2025onlineSpacebounded}.

\medskip
\noindent
The {\em unnormalized conditional} probability is defined as usual, 
\[
  \tMon(a \mid x_1, b_1, \ldots, x_n) = \frac {\Mon(x_1, b_1, \ldots, x_n, a)} {\Mon(x_1, b_1, \ldots, x_n)}.
\]
A tilde is used to remind that this version is unnormalized. 
The following proposition obtains a tight bound, already for this variant. 

Let $h_n = x_1, f(x_1), \ldots, x_{n-1}, f(x_{n-1})$ in the following 2 results. 

\newcommand{\propOnline}{
  If $f$ is computed by a program of prefix-complexity~$k$, then for each sequence $x_1, x_2, \ldots \in \{0,1\}^*$,
  \[
    \sum_{n=1}^\infty 1 - \tMon(f(x_n) \mid h_n, x_n) \,\le\, k\ln 2 + O(1).
  \]
 }
\begin{lemma}\label{prop:errorOnlineClassification}
  \propOnline
\end{lemma}

\noindent
The proof follows the standard one and is given below. 
By corollary~\ref{cor:classifiers_lower} this implies that errors converge faster using online algorithmic probability. 

\begin{corollary}\label{cor:classifiers_difference}
  There exists a computable classifier $f$ and infinitely many $N$ for which there exist strings $x_1, x_2, \ldots, x_N \in \{0,1\}^*$ such that 
  \[
    \lim_N \sum_{n=1}^N 1 - M(f(x_n) \mid  h_n, x_n) = +\infty 
    \qquad \text{ and } \qquad  
    \sum_{n=1}^N 1 - \tMon (f(x_n) \mid h_n, x_n) \le O(1). 
  \]
  where $h_n = [x_1, f(x_1), \ldots, x_{n-1}, f(x_{n-1})]$.  
\end{corollary}

\begin{proof}[Proof of lemma~\ref{prop:errorOnlineClassification}.]
  The proof follows~\cite[theorem 3.8.9]{Hutter2024book}. Since $1-a \le \ln \tfrac 1 a$ for all $a > 0$, 
  \[
      \sum_{n=1}^\infty 1 - \tMon(f(x_n) \mid h_n, x_n) 
     \le \sum_{n=1}^\infty \ln \frac 1 {\tMon(f(x_n) \mid h_n, x_n)}. 
   \]
   Expanding the definition of conditional online complexity, 
   \[
     \le \ln \Big(\frac {\Mon(x_1)} {\Mon(x_1, f(x_1))} \cdot \frac {\Mon(x_1, f(x_1), x_2)} {\Mon(x_1, f(x_1), x_2, f(x_2))}
		  \ldots \cdot \frac  {\Mon(h_n, x_n)} {\Mon(h_n, x_n, f(x_n))}\Big).
    \]
   By definition of online semimeasure, each denominator cancels with the next numerator. 
   Thus this equals 
   \[
     \ln \frac  {\Mon(\textnormal{empty list})} {\Mon(h_\infty)} \le \ln \frac 1 {\Mon(h_\infty)}. 
   \]

   \medskip
   \noindent
   Now it remains to show that $-\log \Mon(h_\infty) \le \K(q) + O(1)$ where $q$ is a program that computes~$f$ on some (plain) reference machine~$V$. 
   As usual, we do this by constructing an online semimeasure $\Pon$ and invoking optimality from proposition~\ref{prop:algorithmicOnlineProbability}. 
   Let $U$ be the optimal prefix machine in the definition of $\K$ and let 
   \[
     \Pon(x_1, b_1, \ldots, x_n, b_n) = \sum \{2^{-|p|} : V(U(p), x_i) = b_i \text{ with } i = 1, \ldots, n\}.
   \]
   Since $U$ is prefix-free, one has $\Pon(\textnormal{empty list}) \le 1$. 
   For lists ending with $x_n$, let $\Pon$ be defined by dropping $x_n$.  
   By construction, $\Pon(h_\infty) \ge 2^{-k}$ if $f$ is computed a program of prefix-complexity $k$. 
   By definition of $\Mon$, one has $\Mon \ge \Omega(\Pon)$, thus 
   \[
     \log \tfrac 1 {\Mon(h_\infty)} \le \log \tfrac 1 {\Pon(h_\infty)} + O(1) \le k + O(1).
   \]
   The proposition is proven.  
\end{proof}

\end{document}